\documentclass{WileyMSP-template}
\usepackage{xcolor}
\usepackage[utf8]{inputenc}
	
\usepackage{soul}

\begin{document}

\pagestyle{fancy}
\lhead{L. Gierster and O. Turkina et al., Adv. Sci. (2024) http://doi.org/10.1002/advs.202403765} \rhead{\includegraphics[width=2.5cm]{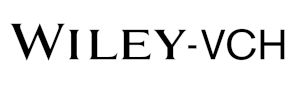}}

\title{Right On Time: \\Ultrafast Charge Separation Before Hybrid Exciton Formation}

\maketitle


\author{Lukas Gierster*},
\author{Olga Turkina*},
\author{Jan-Christoph Deinert},
\author{Sesha Vempati},
\author{Elsie Baeta},
\author{Yves Garmshausen},
\author{Stefan Hecht},
\author{Claudia Draxl},
\author{Julia Stähler}



\begin{affiliations}

Dr. L. Gierster, Dr. Y. Garmshausen, Prof. S. Hecht, Prof. J. St\"ahler \\
Department of Chemistry,  Humboldt-Universit\"at zu Berlin,
Brook-Taylor-Str. 2, 12489 Berlin, Germany\\

Dr. L. Gierster, Dr. J.-C. Deinert, Prof. S. Vempati, E. Baeta, Prof. J. St\"ahler \\
Department of Physical Chemistry, Fritz-Haber-Institut der
Max-Planck-Gesellschaft, Faradayweg 4-6, 14195 Berlin, Germany\\

O. Turkina, Prof. C. Draxl\\
Department of Physics,  Humboldt-Universit\"at zu Berlin,
Newtonstr. 15, 12489 Berlin, Germany\\

Prof. S. Hecht, Prof. C. Draxl\\
Center for the Science of Materials Berlin, Humboldt-Universit\"at zu Berlin,
Zum Gro\ss en Windkanal 2, 12489 Berlin, Germany\\

*Email Addresses: lukas.gierster@hu-berlin.de, turkina@physik.hu-berlin.de

\end{affiliations}


\keywords{hybrid system, excitons, charge transfer, ultrafast}

\begin{abstract}
\begin{justifying}

Organic/inorganic hybrid systems offer great potential for novel solar cell design combining the tunability of organic chromophore absorption properties with high charge carrier mobilities of inorganic semiconductors. However, often such material combinations do not show the expected performance: while ZnO, for example, basically exhibits all necessary properties for a successful application in light-harvesting, it was clearly outpaced by TiO$_2$ in terms of charge separation efficiency. The origin of this deficiency has long been debated. 
This study employs femtosecond time-resolved photoelectron spectroscopy and many-body \emph{ab initio} calculations to identify and quantify all elementary steps leading to the suppression of charge separation at an exemplary organic/ZnO interface. We demonstrate that charge separation indeed occurs efficiently on ultrafast (350~fs) timescales, but that electrons are \emph{re}captured at the interface on a 100~ps timescale and subsequently trapped in a strongly bound (0.7~eV) hybrid exciton state with a lifetime exceeding 5~$\mu$s. Thus, initially successful charge separation is followed by \emph{delayed} electron capture at the interface, leading to apparently low charge separation efficiencies. 
This finding provides a sufficiently large timeframe for counter-measures in device design to successfully implement specifically ZnO and, moreover, invites material scientists to revisit charge separation in various kinds of previously discarded hybrid systems.

\end{justifying}
\end{abstract}


\section{Introduction}
\begin{justifying}
Driven by the rising demand for renewable energies, tremendous efforts are made to find cost-effective alternatives for silicon solar cells~\cite{Polman2016}. Hybrid systems consisting of molecules with strong absorbance of sun light and semiconductors with high charge carrier mobilities promise to be a solution, for example Grätzel-type solar cells~\cite{Munoz-Garcia2021, Boucle2012}. TiO$_{2}$ is the most commonly used semiconductor, but ZnO is - among other semiconductors - a widely investigated alternative~\cite{Munoz-Garcia2021, Wibowo2020} and expected to perform even better due to its higher electron mobility and ease of nanostructure fabrication~\cite{Wibowo2020}. However, the charge conversion efficiencies of ZnO-based systems are still lower than those TiO$_{2}$-based ones. Nevertheless, considerable progress has been achieved by doping \cite{Wibowo2020} and core-shell structures, e.g. by combining ZnO with TiO$_{2}$~\cite{Law2006}.

Despite extensive efforts, the origin of the lower charge conversion efficiency of ZnO is still under debate~\cite{Neppl2021}. While, depending on the chromophore, the photoinjection of electrons into TiO$_{2}$ can occur in the sub-ps time domain~\cite{Munoz-Garcia2021}, the injection into ZnO is reported to be delayed and to occur in a two-step process of several hundreds of femtoseconds and few to hundreds of picoseconds~\cite{Strothkamper2013}. This  allows for electron-hole recombination before charge separation occurs,  leading to losses in the solar cell performance \cite{Sobus2014}. For decades, several options are discussed to explain this difference: (i) the lower density of conduction band states in ZnO compared to TiO$_{2}$~\cite{Wei2016}, (ii) long-lived molecular~\cite{Borgwardt2016} or interfacial trap states~\cite{Strothkamper2013}, which both depend on the molecules used, and (iii) hybrid charge transfer excitons or hybrid excitons (HX) for short, i.e. interfacial electron-hole pairs with the electron in the inorganic and the hole in the organic layer, which trap the charges at the interface. In the latter case, the binding forces should be much stronger for ZnO due to the lower dielectric constant compared to TiO$_{2}$ ($\epsilon_\mathrm{ZnO} =8$, $\epsilon_{\mathrm{TiO}_2} =80$~\cite{Munoz-Garcia2021}). This is comparable to all-organic interfaces, where strongly (hundred's of meV) bound, in this community commonly called \emph{charge transfer} (CT) excitons are observed delaying charge separation \cite{Zhu2009,Bakulin2012,Jailaubekov2013,Wang2017}. Hence, many authors favour this scenario to explain the ultrashort dynamics on fs to few ps time scales~\cite{Nemec2010, Vaynzof2012a,Siefermann2014,Neppl2021,Furube2014}. However, systematic variation of molecule length and, thus, the position of the photohole with respect to the interface, shows no dominant effect on the ultrafast dynamics~\cite{Strothkamper2013}. 

HX at organic/ZnO interfaces were indirectly observed in electroluminescence studies that show light emission matching the energy difference of the ZnO CB and the HOMO of the molecule minus a reorganization energy of several hundred meV~\cite{Piersimoni2015,Panda2016,Hormann2018}. They are also reported to affect solar cell performance~\cite{Eyer2017}, as do CT excitons in all-organic solar cells~\cite{Benduhn2017}. Clearly, experimental access to the formation dynamics of hybrid excitons at ZnO interfaces is needed to confirm their impact on the charge separation efficiencies and develop strategies to circumvent their detrimental influence.

A \emph{direct} measurement of hybrid excitons and their formation dynamics is, unfortunately, still missing. The widely used transient absorption spectroscopy is not sensitive to electron localization and relies on optical transition dipole moments. Beyond optical approaches, however, progress has been made recently: changes to the potential energy surfaces on timescales between 10's of ps up to $\mu$s were attributed to the formation of localised electrons in ZnO within $<$6~nm from the interface using time-resolved X-ray photoelectron spectroscopy~\cite{Neppl2021}. These results are in line with the observation of defect excitons at ZnO surfaces~\cite{Gierster2022} and ultrafast THz spectroscopy, which detects electrons retained at the interface for times exceeding the ps range after ultrafast photoinjection~\cite{Nemec2010}. Also, time-resolved two-photon photoelectron (TR-2PPE) spectroscopy has been used to study dynamic electron localization at semiconductor surfaces~\cite{Borgwardt2019}, organic/organic~\cite{Jailaubekov2013,Wang2017} and inorganic/organic interfaces~\cite{Gundlach2006,Strothkamper2013}. Electron movement from surface to bulk or vice versa is detected as a decrease/increase of the electron intensity due to the surface sensitivity of 2PPE~\cite{Borgwardt2019,Gundlach2006,Wang2017}. An unambiguous assignment, including knowledge of the formation pathway, (lateral) localization, and binding energy of hybrid charge transfer electrons is, however, still lacking. 

It is important to note that - beyond the bare assignment of the culprit of inefficient charge separation - knowledge of the elementary steps in the formation process of hybrid excitons is indispensable: if electron and hole are trapped at the interface immediately after photoexcitation, the carriers can hardly be separated and are lost for light-harvesting. However, if hybrid exciton formation proceeds via intermediate excited states with sufficiently long lifetimes, strategies to prevent interfacial trapping could be developed.



In this work, we disentangle the charge injection and hybrid exciton formation at an organic/inorganic interface using TR-2PPE combined with \emph{ab initio} many-body perturbation theory, employing the Bethe Salpeter equation (BSE). The former has been successfully used in the past to elucidate exciton and carrier dynamics at metal-organic, semiconductor-organic, and organic-organic interfaces as well as in metals and inorganic semiconductors.\cite{Strothkamper2013,Petek1997,Petek2014,FieldingThornton2019,Hoefer2022,Aeschlimann2020,Tegeder2012} To enable comparison between experiment and theory, we use a well-characterized~\cite{Vempati2019} model system consisting of a single crystal ZnO(10-10) surface and a tailored organic molecule, \emph{p}-quinquephenyl-pyridine (5P-Py, see \textbf{Figure~\ref{fig1}a)}. We isolate the impact of electronic coupling and electron-hole Coulomb interaction and observe sub-ps charge injection dynamics by following the non-equilibrium population dynamics of the lowest unoccupied molecular orbital (LUMO) using two distinct excitation pathways, \emph{interfacial} and \emph{intramolecular}, as illustrated in \textbf{Figure~\ref{fig1}b,c}. 
Based on this, we can rule out that (i) density of state effects and (ii) molecular or interfacial trap states hinder electron injection. On the contrary, after the electrons transfer to ZnO on femtosecond timescales, they escape from the probed volume into the bulk of the semiconductor. Remarkably, however, the electrons \emph{return} to the interface within 100(50)~ps. They populate an electronic state at 0.71(5)-0.91(5)~eV below the conduction band (CB) minimum, in good agreement with our BSE calculations that yield a binding energy of 0.7~eV  for hybrid excitons localised at a nanometer distance from the photohole.  This electron recapture at the interface is driven by the attractive Coulomb potential of the hole on the molecule and does not occur for interfacial excitation, allowing the solid assignment to a hybrid exciton state. The HX lifetime exceeds the inverse laser repetition rate of 5~$\mu$s, leading to a photostationary state in our experiments. Our study, thus, reveals the complete charge transfer sequence at a model organic-inorganic interface and resolves the long-standing issue of low ZnO performance in hybrid solar cells. Not only is ultrafast electron injection into ZnO clearly possible at rates comparable to record values reported for TiO$_{2}$~\cite{Ernstorfer2006, Li2006}; we find that HX formation proceeds by \emph{delayed} attraction of electrons to the interface. This finding, firstly, suggests that efficient hybrid solar cells can be designed using ZnO if interfacial trapping is avoided by funneling either photoholes or -electrons away from the interface within the hybrid exciton formation time of 100(50)~ps. Moreover and more generally, however, the discovery of a sufficiently large window of opportunity for charge extraction strongly encourages reconsideration of materials that were previously discarded for light-harvesting applications due to insufficient charge separation efficiencies.

\begin{figure}
\includegraphics{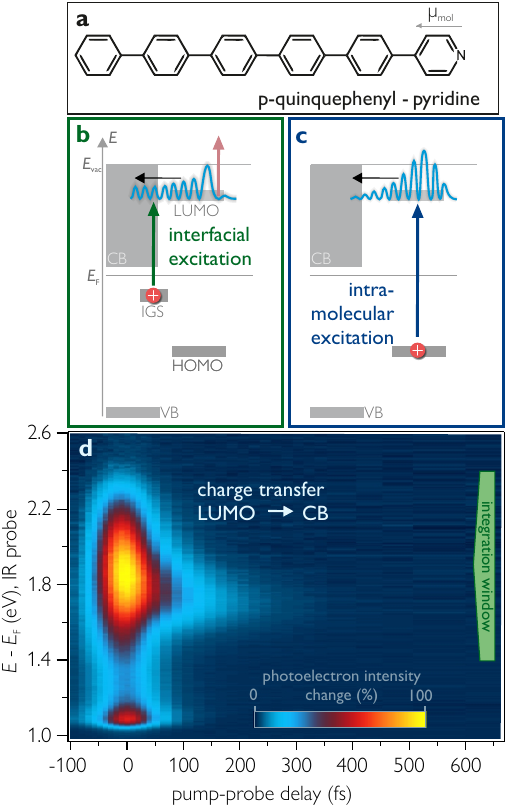}
\centering
\caption{a) The 5P-Py molecule. b,c) Schematics of interfacial and intramolecular excitation. d) Time-resolved PE intensity change (false colors)  after interfacial excitation with h$\nu_\mathrm{pump}=2.52$~eV,  h$\nu_\mathrm{probe}=1.53$~eV as a function of intermediate state energy (left axis) and pump-probe time delay (bottom) after subtraction of the negative delay background.  The peak at 1.8~eV results from the transient population of the 5P-Py LUMO as assigned previously~\cite{Vempati2019}. The signal at lower energies stems from secondary electrons near the low-energy cut-off.}
\label{fig1}
\end{figure}

\begin{figure}
\includegraphics{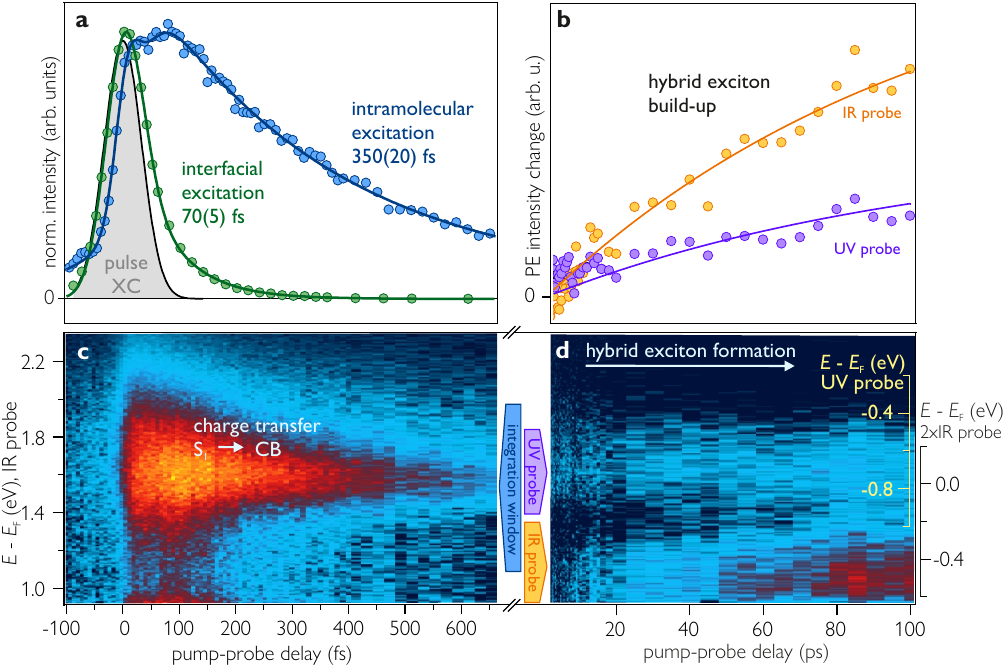}
\centering
\caption{a) Time-dependent PE signal  for interfacial (green) and intramolecular excitation (blue) extracted from Figure~\ref{fig1}d and Figure~\ref{fig2}c, respectively, and laser pulses' envelope (gray). b) Hybrid exciton population build-up extracted from Figure~\ref{fig2}d including fit of the pump-probe delay-dependent changes to photostationary state intensities. c) Time-resolved PE intensity change (false colors) after intramolecular excitation with h$\nu_\mathrm{pump}=3.9$~eV. The signal at 1.6~eV stems from the transient $S_1$ population as assigned previously~\cite{Vempati2019}. d) Build-up of hybrid exciton population probed via changes to the photostationary state intensity when using UV (yellow energy axis) and IR photons (black energy axis, right).For details see \textbf{Section 2.3}. In c,d the negative delay background has been subtracted.}
\label{fig2}
\end{figure}


\section{Results}
\subsection{Strong coupling and ultrafast charge transfer}
As mentioned above, we characterized the energy level alignment at the 5P-Py/ZnO(10-10) interface in a previous publication~\cite{Vempati2019}. Briefly, as determined by photon energy-dependent 2PPE measurements, the system exhibits two spectral features in the vicinity of the Fermi energy: a spectrally broad, occupied in-gap state (IGS) just below the Fermi energy that is localized at the interface and a normally unoccupied level, which is assigned to the 5P-Py LUMO or $S_1$, depending on the population mechanism. Both are illustrated in the energy level diagram in Figure~\ref{fig1}b.

In order to elucidate the degree of electronic coupling between excited molecular levels and the ZnO CB, we measure a non-equilibrium population decay from excited states in the molecule into the ZnO. To do so, we use two distinct excitation pathways in the following. In the case of \emph{interfacial} excitation, we transiently populate the 5P-Py LUMO by resonant excitation from the interfacial IGS (Figure~\ref{fig1}b) using h$\nu_\mathrm{pump}= 2.52$~ eV, equal to the previously determined energy difference between IGS and LUMO. This leads to an excited electron population in the LUMO without creating a photohole in the molecular layer. For \emph{intramolecular} excitation, we drive the $S_0\rightarrow S_1$ transition at h$\nu_\mathrm{pump}= 3.9$~ eV (Figure~\ref{fig1}c), which we determined by photoluminescence excitation (PLE) spectroscopy~\cite{Vempati2019}.

Figure~\ref{fig1}d shows the time-dependent photoelectron intensity change as a function of energy with respect to $E_\mathrm{F}$ (left axis) and pump-probe time delay $\Delta t$ (bottom axis) after \emph{interfacial excitation} (h$\nu_\mathrm{pump}= 2.52$~ eV). At temporal overlap ($\Delta t = 0$) of pump and probe (h$\nu_\mathrm{probe} = 1.53$~eV), an intense signal is observed in the 5P-Py LUMO. It peaks at 1.86 eV and has a full width at half maximum of 0.5 eV in agreement with our previous work~\cite{Vempati2019}. The non-equilibrium population of the LUMO fully decays within few hundred fs. The band integral across the peak yields the green cross correlation (XC) trace in \textbf{Figure~\ref{fig2}a}, which can be accurately fitted with a single exponential decay with $\tau_\mathrm{interface} = 70(5)$~fs convolved with the laser pulses' envelope (gray in Figure~\ref{fig2}a, see \textbf{Section~\ref{PE}} for details). This population decay of the LUMO is much faster than typical electron-hole recombination times in organic molecules~\cite{Dimitrov2016, Faulques2002} and attributed to charge transfer into the ZnO CB. We conclude that the LUMO is strongly coupled to the CB and that low charge separation efficiencies cannot result from a low density of states in ZnO, as efficient charge transfer is evidently possible.

We now address the impact of a photohole on the charge transfer dynamics by resonantly exciting the 5P-Py molecules at their absorption maximum~\cite{Vempati2019}. In order to avoid bulk effects of the organic film which affect the lifetime of excited states in the molecules \cite{Hoefer2022,Aeschlimann2020,Tegeder2012}, we ensure monolayer coverage (see Section~\ref{prep} for details). With the \emph{intramolecular} excitation, the electron wave function is expected to be more localised on the molecule (cf. schematic in Figure~\ref{fig1}c) thereby reducing wave function overlap with the ZnO and, consequently, reducing the transfer rate.  Figure~\ref{fig2}c shows the excited state population dynamics after \emph{intramolecular} excitation, lasting  considerably longer than upon \emph{interfacial} excitation. For quantitative comparison, we take the band integral across the peak (blue energy interval in Figure~\ref{fig2}c). Its temporal evolution is shown in Figure~\ref{fig2}a (blue). As for interfacial excitation, a single exponential population decay fits the data (see Section~\ref{fitting} for more details). The time constant $\tau_\mathrm{molecule} = 350(20)$~fs is five times larger than $\tau_\mathrm{interface}$.

When comparing the ultrafast charge transfer for \emph{interfacial} and \emph{intramolecular} excitation, the position of the photohole needs to be considered. Localized on the 5P-Py molecule in the case of \emph{intramolecular} excitation, the resulting Coulomb potential leads to a stronger localization of the electron wave function on the molecule. As a consequence, the electron has a reduced wave function overlap with the ZnO, which decreases the electronic coupling and, thus, increases the lifetime of the excited state. It should be noted, however, that the charge transfer is (with a time constant of 350~fs) still ultrafast and, hence, the electronic coupling is still strong. We conclude that the influence of the hole on the ultrafast photoinjection is minor, in agreement with reference~\cite{Strothkamper2013}, and that the reduced charge conversion efficiencies at ZnO interfaces must have a different origin.


\subsection{Strongly bound, long-lived hybrid excitons}

To elucidate the cause for suppressed charge separation, we performed {\it ab initio} calculations of $n$P-Py/ZnO ($n=0,1,2$ phenyl rings attached to pyridine) interfaces. Our calculations show that the HOMOs are quite localised on $n$P-Py as illustrated by the charge density plots in \textbf{Figure~\ref{fig3}a}. They do, however, weakly extend into the ZnO. BSE calculations (see Supplementary Material (SM) and reference~\cite{Turkina2019} for details) show that a hybrid exciton can be formed with a binding energy of 0.2~eV. Figure~\ref{fig2}b shows the quasiparticle band structure of Py/ZnO (the corresponding results for $n=1,2$ can be found in the SM) and illustrates by the color code the respective contribution of the electron (light blue) and hole (light red) to the HX wave function. Note that the hole is in the HOMO of the molecule.

\begin{figure}
\includegraphics{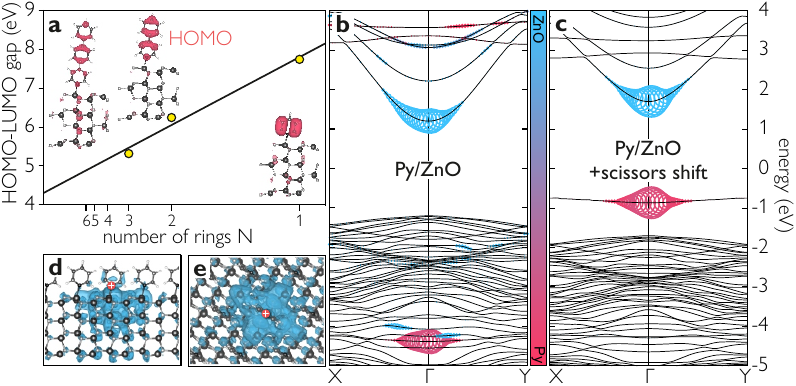}
\centering
\caption{a) HOMO-LUMO gap and HOMO electron density of $\emph{n}$P-Py as calculated by DFT with the number of rings including the pyridine unit $N=n+1$. \textbf{b)} Band structure of Py/ZnO(10-10) as calculated by DFT+$GW$ and hybrid exciton contributions based on BSE. c) Analogous calculations for the HOMO scissor-shifted into the band gap as observed in the experiment and adjusted ZnO band gap. d,e) Side and top view of the electron distribution of the exciton wave function, with the hole localised at the molecule (red dots). }\label{fig3}
\end{figure}

Unfortunately, a full many-body treatment consisting of the $GW$ approximation and a subsequent BSE calculation for the experimentally investigated 5P-Py molecule is computationally not feasible. We therefore use a simplified approach to simulate the HX at the 5P-Py/ZnO interface. Thereby, we make use of the fact that the band gap of $n$P-Py shrinks linearly with the inverse number of rings (cf. Figure~\ref{fig3}a) as commonly known for such conjugated systems. Experimentally, we determined the position of the LUMO 1.8~eV above the Fermi energy, i.e. 1.6~eV above the ZnO CB minimum, and the optical gap to be 3.9~eV~\cite{Vempati2019}. ZnO exhibits a wide band gap of 3.4~eV~\cite{Deinert2015}. Accordingly, we adjust the HOMO-LUMO distance and the ZnO band gap by scissors shifts applied to the DFT bands of Py/ZnO. The resulting band structure is displayed in Figure~\ref{fig3}c. This "trick" is justified, as the character of the bands involved in the HX are the same in all the systems. The BSE calculation based on these bands yields the HX with an increased binding energy of 0.7~eV (0.2~eV without scissors shift, see above) due to the stronger localization of the photohole on the molecule, since the HOMO is now residing in the ZnO band gap. Figures~\ref{fig3}d and e show the electron density distribution (40$\%$ equal density surface) of the HX wave function with the hole position fixed on the molecule. Clearly, the electron is residing in the ZnO. Integration of the charge density (cf. SM) shows that 78$\%$ of the electron density resides in the first few nm of the ZnO with a lateral full width at half maximum of 1.4~nm.

Note that the large HX binding energy suggests that the electronic level of the hybrid exciton, which would be probed in photoemission, lies significantly below the conduction band minimum of ZnO (cf. Figure~\ref{fig1}c), which is located at only 0.2~eV above the Fermi energy $E_\mathrm{F}$~\cite{Ozawa2003}. The HX level, would, thus, lie significantly below $E_\mathrm{F}$, i.e. not only within the band gap of ZnO, but also with all adjacent defect and/or polaron states being occupied. Such energy level alignment would quench all potential HX decay channels except for recombination, as all available electronic states (even within several 100~meV) are occupied below $E_\mathrm{F}$. This would inevitably lead to long lifetimes in the order of 100's of ps \cite{Deinert2014, Gierster2021}, ns~\cite{Foglia2019}, or even up to $\mu$s~\cite{Gierster2022}, possibly exceeding the inverse repetition rate of our laser system (5~$\mu$s), such that a photostationary population could be detected even in single-color \mbox{photoemission} experiments~\cite{Gierster2022} (details will be discussed below). In order to probe such metastable states at low energies, either (UV) photon energies exceeding the work function $\Phi=E_\mathrm{vac}-E_\mathrm{F}$ are needed, or multiphoton photoemission with IR or VIS photons is required as illustrated in the inset of \textbf{Figure~\ref{fig4}a}.

The purple spectrum in Figure~\ref{fig4}a shows the single-color (h$\nu_\mathrm{UV}=3.90$~eV) 1PPE intensity distribution as a function of final state energy. Beyond the secondary electron tail that is cut off at 2.4~eV,  indicating the low work function of the 5P-Py sample, the spectrum exhibits a broad peak at $E_\mathrm{final}-E_\mathrm{F}=3.19(5)$~eV, corresponding to an initial state energy of 0.71(5)~eV below $E_\mathrm{F}$ (cf. gray energy axis). The same signature is also detected in the orange single-color (h$\nu_\mathrm{IR}=1.55$~eV) 2PPE spectrum in Figure~\ref{fig4}a. It is located at lower final state energy due to the photon energy difference $\Delta\mathrm{h}\nu=\mathrm{h}\nu_\mathrm{UV}-2\mathrm{h}\nu_\mathrm{IR}$ and, therefore, superimposed with the secondary electron tail. With respect to the bulk conduction band minimum of our ZnO sample, which lies 0.2~eV above $E_\mathrm{F}$~\cite{Deinert2015}, the state has a binding energy of 0.91(5)~eV, no matter whether it is probed by UV or IR light. However, downward surface band bending at the surface of ZnO(10-10) up to $E_\mathrm{F}$ was observed before~\cite{Deinert2015, Gierster2022} and is highly likely also for the interface with 5P-Py, not least because of the strongly reduced work function. Based on this, we deduce an experimental binding energy of 0.71(5)-0.91(5)~eV. Although a quantitative agreement with our theory, which only takes into account significantly smaller molecules, cannot be expected \emph{a priori}, the excellent match of this binding energy with our calculations should be noted. The impact of the size of the molecule on the binding energy is either small (for example due to hole localization at the interface due to the Coulomb potential of the electron in the ZnO) or is balanced by other factors like polaronic distortions and defects.

\begin{figure}
\includegraphics{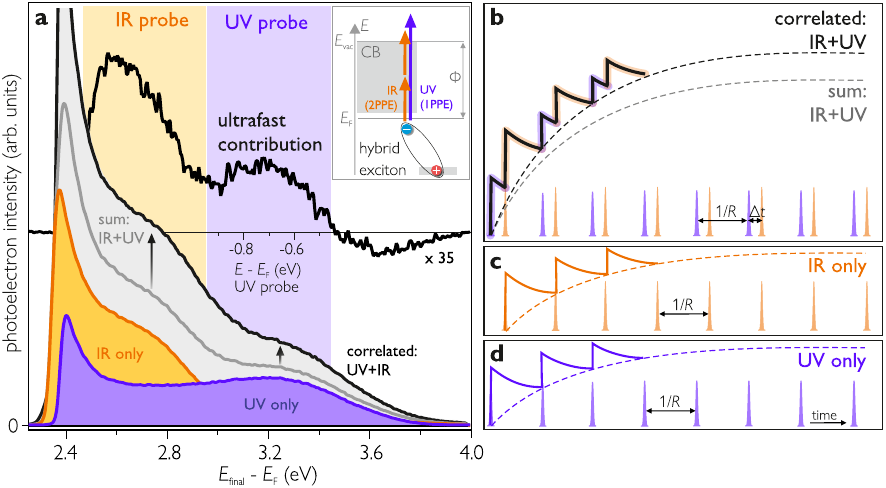}
\centering
\caption{Long-lived hybrid excitons as photostationary states in PE spectra. a) Comparison of single color PE spectra (purple, orange) and their sum (gray) with correlated signals (black). Difference spectrum for $\Delta t=+85$~ps and -3~ps is offset and upscaled for clarity. Inset: The hybrid exciton is probed by IR and UV photons in a 2PPE and 1PPE process, respectively. b-d) Build-up of photostationary states for (d) UV and (c) IR illumination only, as well as for two-color experiments.}
\label{fig4}
\end{figure}

To test the hypothesis that the state observed in the single-color spectra is resulting from a  photostationary population, we inspect the nature of this state by comparison of the sum of the single color spectra (gray line) with a two-color PE spectrum at negative pump-probe time delays (black line). If the electronic state was originating from an ordinary occupied state, no correlation of the laser pulses should be observed and the negative delay spectrum should coincide with the sum of the single-color spectra. Clearly, this is not the case: Both, the UV- and IR-probed peaks, have larger intensities in the black spectrum, when both colors are used in the experiment, demonstrating a \emph{correlated} signal of a state below $E_\mathrm{F}$ at negative pump-probe delays.

This observation can be understood when considering an excited state lifetime exceeding the inverse repetition rate of the laser system ($1/R=5$~$\mu$s). In single-color PE, each consecutive laser pulse populates and depopulates the long-lived species, leading to the build-up of a  photostationary state with an intensity that depends on the (de)population probabilities as illustrated by Figure~\ref{fig4}c,d. When both colors are used, separated e.g. by a pump-probe delay $\Delta t>0$ (cf. Figure~\ref{fig4}b), the excited state population generated by the UV may be probed by IR photons and vice versa. A \emph{correlated} photostationary state (PSS) is detected (black) that has a different intensity than the mere sum of the single-color signals (gray).

Based on the above, we conclude that \emph{intramolecular} excitation is not only followed by ultrafast electron transfer from the 5P-Py to the ZnO (cf. \textbf{Section 2.1}), but also causes the formation of a photostationary state with a lifetime exceeding the inverse repetion rate of our laser system, larger than 5~$\mu$s. This is rationalized by the experimentally observed large binding energy of 0.71(5)-0.91(5)~eV with respect to the ZnO CB continuum, which shifts the photostationary state below the Fermi energy where all decay channels except for recombination are blocked. These findings are in line with our theoretical result that strongly bound hybrid excitons with a binding energy of 0.7~eV can form at \emph{n}P-Py/ZnO interfaces.

\subsection{Formation dynamics of the hybrid excitons: ultrafast, but delayed}

It is tempting to assume that a photo\emph{stationary} state does not exhibit any dynamics once it is built-up. However, a PSS is only stationary with respect to the fixed pump-probe delay $\Delta t$ at which the spectrum is taken. In other words, for different $\Delta t$, different PSS may form, for instance because the succeeding pulse interferes with the formation dynamics of the transient species launched by the preceding laser pulse as outlined in detail in the SM and in reference~\cite{Lukas_PhD}. Subtraction of photostationary spectra at different $\Delta t$ yields such dependence of the PSS intensity on the pump-probe time delay.

The black, unfilled spectrum in Figure~\ref{fig4}a is a representative difference spectrum generated by subtracting $\Delta t=-3$~ps data from the spectrum at $+85$~ps. It is shifted upwards for clarity and significantly upscaled (x35). The spectrum illustrates that the average PSS population is only slightly, but notably higher if the UV photons are absorbed 85~ps \emph{before} the IR pulse arrives than if the UV pulse succeeds the VIS pulse by 3~ps. In other words, the photostationary population is higher for a pulse train with $\Delta t=85$~ps than for $\Delta t=-3$~ps. Based on this, we conclude that intramolecular excitation of 5P-Py leads, in addition to the observation of a PSS to \emph{delay-dependent} photostationary signals.

Having demonstrated that the PSS differs for different pump-probe time delays $\Delta t$, we return to the time-resolved data after intramolecular excitation.  Figure~\ref{fig2}d shows the time-dependent change of the PE intensity for pump-probe delays up to 100 ps. The first important observation is that the PE intensity is almost in the noise floor for the first few ps, which means that the complete 5P-Py S$_1$ population has decayed by electron transfer to the ZnO as discussed above. The lack of PE intensity can either result from (i) decay to the ground state, (ii) transfer to bands at large $k$-vectors inaccessible by our probe photon energies, or (iii) transfer to the bulk of ZnO such that the mean free path of the photoelectrons does not allow photoemission. However, as a PSS is formed, at least part of the system cannot have relaxed to the ground state. Further, previous experimental~\cite{Deinert2014} and theoretical~\cite{Zhukov2010} studies showed that hot electrons relax to the CB minimum of ZnO at the $\Gamma$-Point on sub-ps timescales. We can, thus, conclude that the vanishing of the photoelectron intensity is resulting from (iii) transfer to the bulk of ZnO.

After ca.~10~ps, a slow intensity build-up is observed in Figure~\ref{fig2}d, at the initial state energy of the photostationary state introduced in \textbf{Section 2.2}. As discussed above, it appears in a double peak structure due to probing by photoemission with IR \emph{and} UV photons (yellow and black axes on the right), respectively, which is possible due to the long lifetime of this species. Clearly, the intensity of both peaks depends on the pump-probe time delay and rises gradually. This increase of PE intensity indicates that electrons return to the interface, where they can again be probed by PE.  In order to quantify the rise time of the signal, i.e. the electron recapture time constant, we extract band integrals of both peaks and plot the transients in Figure~\ref{fig2}b. A global single exponential fit (black lines) yields a rise time of 100(50)~ps.

We conclude from these experimental findings that the $S_1$ electron population created by intramolecular excitation first transfers into the bulk of the ZnO and then returns to the hybrid interface on much longer picosecond timescales. The resulting feature in photoemission is a photostationary signal, which is caused by the microsecond lifetime of the state. While all of these experimental observations are consistent with our calculations, which suggest that this state is the signature of a hybrid exciton, it is important to consider alternative interpretations before drawing conclusions.

No matter which type of trap state is responsible for the photostationary state, the experimental observation of the electron \emph{recapture} at the interface demonstrates that an attractive driving force must be present. The potential energy landscape must, thus, exhibit a minimum in the interface region. Such potential minimum could be created by the attractive Coulomb potential of the photohole on the 5P-Py molecule as in the HX scenario suggested by theory. Beyond hybrid excitons, however, organic/inorganic interfaces and metal oxide surfaces are known to exhibit defect and polaron states that are long-lived and can trap charge carriers efficiently~\cite{Rajh1999, Wang2003, FieldingThornton2019,Borgwardt2019,Gierster2022,Sezen2015}. In order to provide an attractive interfacial trap for electrons that already relaxed to the CB minimum of ZnO, these states must lie at lower energies than the CB minimum on the one hand and must be unoccupied, i.e. lie above the Fermi energy, on the other hand. Due to the strong \emph{n}-type doping of our ZnO sample, this leaves a narrow energy window of up to 200~meV above $E_\mathrm{F}$. In this scenario, the trap state would have to have a strong polaronic character in order to accommodate the large binding energy of at least additional 500~meV \emph{below} $E_\mathrm{F}$.

While our previous studies on the electronic structure and dynamics of the ZnO(10-10) surface~\cite{Deinert2015, Deinert2014, Gierster2021, Gierster2022, Staehler2017} show no indication of defect and/or electron polaron states \emph{above} $E_\mathrm{F}$ and no build-up of electron density on a 100~ps timescale as observed here, the trap states could originate from the interface formation with the 5P-Py molecules. In order to disentangle long-lived PSS caused by defects/polarons and hybrid excitons, we make use of the fact that hybrid excitons can only be observed in the presence of a photohole in the molecular layer. Thus, if the PSS were resulting from defect or polaron states, it should also appear in the case of interfacial excitation.

\textbf{Figure~\ref{fig5}} shows a comparison of single color pump (h$\nu_\mathrm{pump}=2.52$~eV, green) and probe (h$\nu_\mathrm{probe}=1.53$~eV, orange) spectra as well as their sum (red) with a spectrum taken at negative time delays (gray), i.e. when the probe pulse precedes the pump. Analogous to the discussion of Figure~\ref{fig4}a in Section~2.2, a photostationary signal would be manifested by a difference between the sum of the single color spectra and the negative delay spectrum. Clearly, this is not the case for interfacial excitation, as the red and the gray spectrum coincide. We conclude that interfacial excitation does not cause a PSS. This observation demonstrates that a photohole on the 5P-Py is needed to create a photostationary signal and, thus the electron recapture at the hybrid interface. The Coulomb potential of the hole attracts the electron to the surface where it forms a hybrid exciton, in agreement with the theoretical prediction.

\begin{figure}
\includegraphics{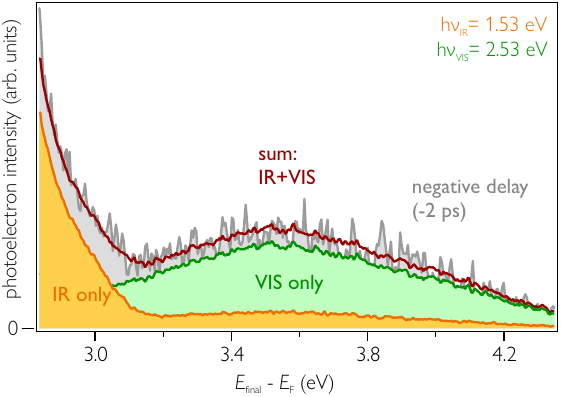}
\centering
\caption{ Comparison of single color PE spectra (green, orange) and their sum (red) with a PE spectrum at negative delays (gray) for interfacial excitation. As the red and the gray spectrum coincide, no photostationary signal is observed. }
\label{fig5}
\end{figure}

\begin{figure}
\includegraphics{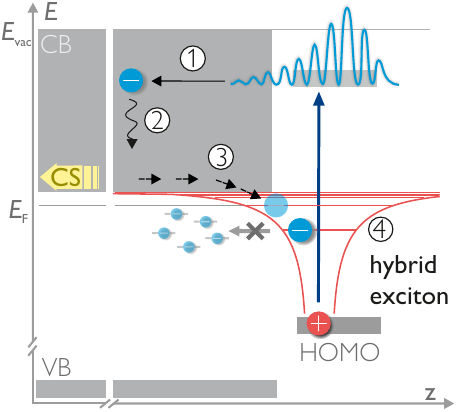}
\centering
\caption{ Scheme for hybrid exciton formation after intramolecular excitation: (1) electron transfer to the bulk of ZnO, (2) relaxation to the CB minimum, (3) electron recapture at the surface to form (4) the hybrid exciton. Adjacent in-gap defect states are below $E_\mathrm{F}$ and, thus, occupied.}
\label{fig6}
\end{figure}

\section{Conclusions}
Based on the above findings, we are able to describe the complete sequence of elementary steps that lead to hybrid exciton formation at organic-ZnO interfaces, illustrated in \textbf{Figure~\ref{fig6}}: Photoexcitation of the chromophore is followed by (1) ultrafast electron transfer to the ZnO, competing with record charge injection times of TiO$_2$~\cite{Ernstorfer2006,Li2006}. This electron transfer is only marginally slowed down by the presence of the photohole on the molecule. The relaxation (2) of the injected electrons is not detected by photoemission, which means that the carriers have left the surface region. It is, however, known~\cite{Zhukov2010,Deinert2014} that electrons reach the minimum of the conduction band quickly on sub-picosecond timescales. Remarkably, only after a substantial delay, the electrons are (3) recaptured at the interface. The hybrid exciton signal builds up with a time constant of 100(50)~ps. This can be rationalized by the long-range Coulomb attraction of the photohole on the chromophore, which is less screened in ZnO than in TiO$_2$. Due to the very large binding energy of the hybrid exciton (4), it takes time for the exciton to reach the lowest exciton state within the potential of the hole, as several 100~meV need to be released upon cooling of the hot exciton. Moreover, except for recombination, all thermally accessible HX decay channels, in particular defect and polaron states, are already occupied, as all electronic levels are filled up to the Fermi energy (cf. Figure~\ref{fig6}). The hybrid exciton can solely decay by recombination. This leads to very long lifetimes beyond 5~$\mu$s. 

We would like to point out that, whilst we do not observe any spectroscopic evidence for additional, unoccupied trap states like defects or polarons that could be occupied by the injected electrons, our findings do not exclude their existence. They could simply be not detectable in our photoemission experiment due to too small transfer matrix elements, for example. Nevertheless, as mentioned in Section~2.3, such states would require a population mechanism above $E_\mathrm{F}$ and a subsequent strong polaronic character to trap the electrons sufficiently deeply to inhibit charge separation.

It is illustrative to compare our findings for the organic/inorganic hybrid system to all-organic photovoltaic systems. The all-organic analogue of the HX are CT excitons, where electrons and holes reside on different entities, but are still experiencing the mutual Coulomb interaction. Charge separation occurs for example, when electrons can escape from the CT state~\cite{Vandewal2014} to the charge separated (CS) state by thermal activation or due to entropy gain in the comparably disordered all-organic system~\cite{Puttisong2018}. Also the delayed formation of a charge transfer exciton ground state is known for organic solar cells \cite{Jailaubekov2013,Wang2017}. However, the trapping occurs much faster (1-2~ps~\cite{Jailaubekov2013, Wang2017} vs. 100~ps as observed here). The main mechanistic difference of our work compared to all-organic systems is that the injected electrons are, after entering the bulk of ZnO (process 2 in Figure~\ref{fig6}) already almost completely separated from the hole and in a \emph{quasi}-CS state with only very weak far-field Coulomb interaction. Inhibition the formation of the lowest-energy hybrid exciton state is therefore expected to be easier than for all-organic systems as discussed in more detail below.

This new and detailed understanding of the delayed formation of hybrid excitons is not only consistent with previous findings in the literature as, for instance, the high binding energy from electroluminescence studies~\cite{Piersimoni2015,Panda2016} or the long lifetime of interfacial trapped electrons in time-resolved XPS~\cite{Neppl2021}. It also directly leads us to concerted strategies to circumvent hybrid exciton formation for enhanced charge separation. The prime insight gained is the finding that hybrid exciton formation occurs in a \emph{delayed} fashion, with a quite substantial build-up time of 100~ps. Knowing that the carriers are actually spatially separated before the electrons are \emph{re}captured at the interface, a comfortable time window is opened up, which allows funneling of photocarriers away from it \emph{before} they bind in the form of stable hybrid excitons: Firstly, the delayed hybrid exciton formation is driven by the attractive Coulomb potential of the hole residing on the molecule. This means that screening of the Coulomb interaction will increase the probability of the electron to escape to the truly charge separated state in which the electron does not experience the Coulomb potential of the hole anymore (cf. CS in Figure~\ref{fig6}). Enhanced screening of the Coulomb interaction reduces the interaction radius and thereby enhances the charge separation efficiency, as confirmed by the use of core-shell concepts (cf. e.g. \cite{Law2006}). Our new understanding of the underlying mechanism will help to improve this strategy further. However, not only the reduction of the attractive Coulomb interaction will be useful, but also concerted doping of the inorganic semiconductor to enhance the screening as, for instance, in  reference~\cite{Wibowo2020}. Lastly, knowledge of the transient separation of electrons and holes unveils that the carriers need to be collected in closer proximity to the organic/inorganic interface: electron capture at the backside of sufficiently thin ZnO films is equally conceivable as scavenging photoholes from the chromophores by using appropriate hole acceptors, or manipulation of the potential energy landscape by built-in electric fields.

Finally, we would like to emphasize that neither hybrid exciton formation nor the above-described strategies to circumvent this undesirable process are constrained to ZnO-based hybrid interfaces. Other semiconductors with comparable dielectric constants (e.g. SnO$_2$ or In$_2$O$_3$~\cite{shi2021}) are very likely to show hybrid exciton formation and, as a result, low charge separation efficiencies as well. Most importantly, the observation of initial, transient charge separation \emph{before} actual trapping occurs is a clear invitation to revisit previously discarded material combinations. The observed low charge separation efficiencies could be a result of delayed trapping as in our exemplary model system. Thus, the above-described strategies to circumvent carrier loss at the interface could indeed prove to be the game changer.

\section{Experimental Section}
 
\threesubsection{Sample preparation}\label{prep}\\
Sample preparation is performed in a UHV chamber with a base pressure below $5\cdot10^{-10}$~mbar. Single crystal ZnO(10-10) samples (MaTecK GmbH) are prepared by Ar$^+$ sputtering (0.75~keV, 10~min) and annealing cycles at 700-900~K for 30 min following established procedures \cite{Diebold2004}. The surface reconstruction is confirmed with LEED. The 5PPY molecule is the aza-analogue of the widely used sexiphenyl, with one nitrogen atom at the end to bind in an ordered fashion at the ZnO surface, for details on the synthesis see reference~\cite{Vempati2019}. 5PPy molecules are evaporated (0.3-0.7~nm/min) from a Knudsen cell onto the freshly prepared ZnO(10-10) surface after carefully degassing of the molecules at 10 K below their evaporation temperature of ca.~530~K. A pre-calibrated quartz crystal microbalance is employed to estimate the thickness of the film. The adsorption of 5PPy on ZnO(10-10) also induces a work function change that is characteristic of the film thickness; a minimum work function of 2.45(5)~eV is reached at 1~ML \cite{Vempati2019}, corresponding to the data shown in Figure \ref{fig2}~c (intramolecular excitation).

\threesubsection{Photoelectron spectroscopy}\label{PE}\\
A scheme of the experimental setup is shown in the SM (Figure S4). After preparation, the sample is transferred \emph{in situ} to the analyzer chamber. Photoelectrons are detected using a hemispherical analyzer (Phoibos 100, SPECS) with an energy resolution of ca. 50~meV. During the photoemission experiments the sample temperature is held at 100~K, and the pressure is kept $<$10$^{-10}$~mbar. A bias voltage $V_\mathrm{bias}$ of -2.5 to -5.5~V is applied to the sample with respect to the analyzer enabling the detection of electrons with zero kinetic energy, which constitute the secondary electron cutoff. The spectra are referenced to the Fermi level $E_\mathrm{F}$ measured at the gold sample holder, which is in electrical contact with the sample. Note that the kinetic energy of $E_\mathrm{F}$ scales with the photon energy h$\nu$, as the measured kinetic energy of electrons originating from $E_\mathrm{F}$ is $E_\mathrm{kin}(E_\mathrm{F}) = \mathrm{h}\nu-\Phi_\mathrm{Analyzer} - eV_\mathrm{bias}$, where $\Phi_\mathrm{Analyzer}$ is the work function of the electron analyzer. The pump and probe laser pulses are generated by a femtosecond 200 kHz laser system (Light Conversion), consisting of a regenerative amplifier (Pharos) feeding two non-collinear optical paramteric amplifiers (Orpheus 2H/3H). Pulse durations are on the order of 30-50 fs. For the intramolecular excitation the pump laser fluence is kept low (0.2 $\mu$J/cm$^2$) due to strong direct photoemission (the pump photon energy of 3.9 eV is larger than the sample work function). The probe laser fluence is 700 $\mu$J/cm$^2$. Due to the high power, 2PPE with the probe laser beam from occupied states is about as efficient as direct photoemission with the pump laser beam.

\threesubsection{Statistical analysis}\label{statistics}\\
The raw time-resolved 2PPE intensity was integrated across the maximum emission angle (14\degree and 7\degree, respectively). In the time-resolved 2PPE spectra presented in Figure~ref{fig1}~d and \ref{fig2}~c,d the photoemission background at negative delay times (-2 ps) was subtracted to show the excited state dynamics. In order to exclude systematic errors the delay range was scanned from negative to positive delays and subsequently from positive to negative delays. The data represents an average of several tens of  scans. The error bars in the fits of the time-resolved data (Figure~\ref{fig2}a,b) represent the standard deviation.

\threesubsection{Data fitting}\label{fitting}\\
The fitting of the photoelectron intensity transients in Figure \ref{fig2}~a is done for interfacial excitation with 

$$f(\Delta t) = A e^{-\Delta t^2} + B e^{-\Delta t/\tau_\mathrm{decay}}$$ 

and for intramolecular excitation with 

$$f(\Delta t) = A e^{-\Delta t^2} + B e^{-\Delta t/\tau_\mathrm{decay}} (1-e^{-\Delta t/\tau_\mathrm{rise}}). $$ 

The first term in both fit functions is a narrow Gaussian function (width 1~fs) to account for two-color 2PPE via virtual states without populating the LUMO from the IGS/the HOMO. A delayed rise ($\tau_\mathrm{rise}=80(10)$~fs) of the LUMO population is observed upon intramolecular excitation, which is likely attributed to internal conversion; two color 2PPE and delayed rise leads to the double peak structure in the data (cf. Figure \ref{fig2}~a). The functions are convoluted with the laser pulses’ cross-correlation measured at the gold sample holder before fitting them to the data. 

\threesubsection{Theory}\label{TH}\\
All calculations have been carried out with the all-electron full-potential code \textbf{exciting}\ \cite{Gulans2014}, based on the frameworks of density-functional theory and many-body perturbation theory. Details are provided in the SM. All input and output files of the calculations can be downloaded from NOMAD \cite{Draxl2019} under the DOI \\ 10.17172/NOMAD/2023.06.14-1.

\medskip
\textbf{Supporting Information} \par 
Supporting Information is available from the Wiley Online Library or from the author.

\medskip
\textbf{Acknowledgements} \par 
We acknowledge funding by the Deutsche Forschungsgemeinschaft (DFG, German Research Foundation) - Project-ID 182087777 - SFB 951.

\medskip
\end{justifying}
%
\bibliographystyle{MSP}
\bibliography{hybrid}




\end{document}